\documentclass[UTF8,a4paper]{article}
\usepackage{amsmath}
\usepackage{bm}
\usepackage{graphicx}
\usepackage{subfigure}
\usepackage{ upgreek }
\usepackage{ stmaryrd }
\usepackage{ mathrsfs }
\usepackage{caption}
\usepackage{epstopdf}
\usepackage{ marvosym }
\usepackage{ tipa }
\usepackage{geometry}
\usepackage{ wasysym }
\usepackage{ulem}
\usepackage{indentfirst}
\usepackage{amsfonts,amssymb,dsfont}
\usepackage{accents}

\usepackage{setspace}
\usepackage{threeparttable}
\usepackage{amsmath,mathtools,amsthm}
\usepackage{array}
\usepackage{extarrows}
\usepackage{upgreek}

\makeatletter
\renewcommand*\env@matrix[1][\arraystretch]{%
	\edef\arraystretch{#1}%
	\hskip -\arraycolsep
	\let\@ifnextchar\new@ifnextchar
	\array{*\c@MaxMatrixCols c}}
\makeatother

\usepackage[pdfstartview=XYZ,
bookmarks=true,
colorlinks=true,
linkcolor=blue,
urlcolor=blue,
citecolor=blue,
pdftex,
bookmarks=true,
linktocpage=true, 
hyperindex=true
]{hyperref}
\usepackage{orcidlink}

\begin{tiny}
	\title{ Ferroelectricity and related effects on carrier transport in type-II Weyl semimetal WTe$_{2}$ thin film}

	\author{
		Chen-Huan Wu
		\orcidlink{0000-0003-1020-5977} 
		\thanks{chenhuanwu1@gmail.com},
\  \
Yujie Ren
		\\
		School of Microelectronics, Southern University of Science and Technology, 518055 Shenzhen, China
	}
\end{tiny}

\begin{document}	\begin{small}
	\maketitle
		\begin{abstract}
			We investigate ferroelectric polariation as well as the formation of long-range order and the carrier density distribution in type-II Weyl semimetal WTe$_{2}$ in $T_{d}$ phase.
			It is been found that
			the metallicity and ferroelectricity can coexist in bulk
			WTe$_{2}$ which has a significant impact on the electrical transport\cite{Sharma}, despite its large conductance.
				Also,
			our theoretical calculation and numerical simulation 
			provide a deeper insight to the electrical structure-dependent dynamics of WTe$_{2}$.
			Base on the two-level approximation
			verify that the polarization stems from uncompensated out-of-plane interband transition of the electrons,
			which is base on the calculations of the dipole transition moment
			(in both the momentum space and frequency domain),
			and we found that the topological character of type-II Weyl system is closely related to the electronic behaviors (like the carrier compensation) and the excitations near the Weyl cone.
			The anisotropy and 
			the topologically protected spin-polarized bulk (Weyl orbit) and surface states in WTe$_{2}$ 
			induce hysteresis, which exhibitspotential in applications of non-volatile energy-efficient data-storage devices.
			Part of the properties of WTe$_{2}$ are also
			shares shared by the thermoelectric properties with other  two-dimensional transition-metal
			dichalcogenides, like the WSe$_{2}$ and MoTe$_{2}$.
		\end{abstract}
		
		{\color{red} $Introduction$}
		As a type-II Weyl semimetal, WTe$_{2}$ has a $Pm$ point group symmetry, and the inversion symmetry is broken.
Such non-centrosymmetricity leads to non-trivial geometric properties for the Bloch states,
and allows higher-order response to external electric field
which contains the components along
more than one principal crystallographic direction\cite{Ortix,Deb,Zhou2}.
		WTe$_{2}$, as a polar metal as well as ferroelectric material,
		contains an intrinsic electrical dipole moment, which does not relies on the external electric field.
		The $dI/dV(V)$ and $dV/dI(I)$ measurements in bulk WTe$_{2}$\cite{Orlova N N,Orlova N N2} show
		sweep direction dependence which indicates that the
		source-drain field variation generates an additional polarization current.
		During the ferroelectrical polarization, 
		the spontaneous polarization vector switchs between
		the two differrent orientations when the oppositely oriented electric field is applied.
		The key experiment for ferroelectricity is the existence of a hysteresis loop
		between polarization and electric field.
		Despite the ferromagnetism in layered T$_{d}$-WTe$_{2}$  can coexist with the Weyl semimetal phase,
		in the absence of doping and defects, highly pure T$_{d}$-WTe$_{2}$ is diamagnetic,
		and the M-H curves show constant linear negative slope in a large scale of magnetic field.
		In the absence of magnetic impurity and defeats,
		the saturation magnetic moment is vanishingly small,
		and the ferromagnetic order is totally overwhelmed by the semimetallic behaviors.
		With broken inversion symmetry and time reversal symmetry,
except the ordinary Hall effect under an external magnetic field, the	anomalous Hall effect can be observed under zero magnetic field and an finite bias current,
like the out-of-plane magnetization which can be generated by the bias current along $a$-axis\cite{Kang}.
				As an orthorhombic noncentrosymmetric material,
		a vertical electric field can break the two-fold screw rotation symmetry of WTe$_{2}$ and generating
		finite Berry curvature dipole which can leads to nonlinear Hall current even in the absence of magnetic field.
		Recent studies also focus on coexistence of ferroelectricity and nonlinear Hall (or anomalous Hall) effects\cite{2},
		which also imples the unique ferroelectric properties and the related carrier transport in WTe$_{2}$.

				{\color{red} $Discussion$}
		Ferroelectricity as an electrical collective phenomenon, is related to the rotation
		of dipoles.
For single layer film the dipole-dipole interaction can induces mirror charges on conducting substrate and results in out-of-plane polarization\cite{Duan C}.
		Such type of ferroelectric polarization in fre layer films
		is realized by interlayer charge transition through the excitations of sliding phonons which carry the electric dipoles,
		and this sliding ferroelectric polarization as well as the 
		shear motion of phonon driven by thermally fluctuating force
		is guaranteed by the 
		high intralayer stiffness\cite{Tang P} up to room temperature.
		The low ferroelectric switching barrier and stable sliding ferroelectric polarization
		make it feasible for the nanoscale polarization manipulation and 
		electric polarization switching modulation.
		Except the effect of shear motion which requires thermal effect,
		the electric dipole moment can be modulated by
		applying of an external electric field.
		Since the ferroelectric polarization will be suppressed
		by the screening of conduction electrons like the metallic gates
		\cite{Sharma P,Tang P},
		and such screening will destroy the electrostatic internal field formed by the
		long-range dipolar order as well as the electrostatic potential
		of ions.
		Also, compared to the few layer films, the bulk crystals are better
		in preserving the long-range dipole order\cite{Tang P}.
		While in the systems with lower dimension,
		the long-range dipole order is suppressed,
		like the few layer 2D crystals or the 1D quantum system where
		the thermalization effect plays a key role\cite{Li W,Wu3}.

		We found that the spontaneous out-of-plane electric
		polarization which can be switched by applying a in-plane electric field,
and a bistability behavior in resistivity is available at room temperature.
The long range electrostatic interaction as well as the dipole order
		reflects the anisotropic characteristic with the long-range Coulomb attraction for parallel
		dipoles along the polarization axis.
		Unlike the point-like Fermi surface in type-I Weyl system,
		there is a Fermi surface containing the electron and hole pockets
		of the near size.
		With the near electron and hole density/mobility,
		there are rich electrical transport phenomenons,
		e.g., in terms of a compensated theory\cite{Pletikosic I}.
		Also, near the contain point of the electron and hole pocktes,
		the out-of plane polarization in WTe$_{2}$ can be observed more easily
		ferroelectrical polarization\cite{Fei Z} and magnetoresistivity.
		Furthermore,
		our theoretical calculation base on the two-level approximation
		verify that the polarization stems from uncompensated out-of-plane interband transition of the electrons,
		which is base on the calculations of the dipole transition moment
		(in both the momentum space and frequency domain),
		and we found that the topological character of type-II Weyl system is closely related to the electronic behaviors (like the carrier compensation) and the excitations near the Weyl cone.
		Though there are states from both layers in the
		conduction and valence bands, the cumulative effect of filled
		states up to the Fermi level results in a net polarization due to an
		imbalance of layer character, particularly near the band edges. The magnitude of this imbalance is larger
		for valence band states compared to the conduction band.

		A slow relaxation process and the dependence on sign of current change was previously 
		reported at room temperature\cite{34}. 
		This conclusion is also confirmed by gate voltage dependencies, so our results can be understood as a
		direct demonstration of the ferroelectric behavior of WTe$_{2}$ in charge transport experiment.
For the ferroelectric polarization of thin film WTe$_{2}$,
the DHM measurement is applied to study by gate voltage dependence
the polarization reversal.
		Similar to the above mentioned experimental results,
		the second-order effect here is still depends on the certain crystal axis due to the
		in-plane anisotropy of WTe$_{2}$.

{\color{red} $Experimental\ results$}
Different to the semiconductor ferroelectricity,
the effect of band bending caused by polarized bound charges in channel can be ignored,
and instead, we have to consider the effect of mobile charges,
which play a essential role in ferroelectric performance
and exhibit difference in large and low current cases.
We found that, at low current ($<1\mu A$),
the polarization is quite small, $\sim 0.0035$ $\mu C/cm^2$ for ten-layer WTe$_{2}$,
and the current is mainly dominated by the polarization current but the leakage current
has a higher sensitivity on the voltage as well as the sweeping direction.

The interlayer charge transfer of mobile carriers has an impact on the out-of-plane polarization up to 
10-20 layers,
where the electric field-switchable polarization could be as large as 1 $\mu C/cm^2$
for maximal current $\sim 100\ \mu A$ and channel length up to $40\mu m$.
Polarization states in conductivity as well as the large polarization is more easy to
realized with small current.
Under small current,
the polarization decrease more rapidly with the decreasing current
than that under high current along the channel,
and the two polarization states can be revealed by the condictivity.
This prove the switching capability of WTe$_{2}$ up to dozens of layers.

We found that,
for current (or conductivity) nonzero in both the positive and negative voltages,
the P-V loop exhibit circular shape,
which is similar to the high frequency result of 10-nm thick HZO film\cite{Zhaomeng},
and the value of polarization is relatively large.
In both cases,
the two polarization states can be observed in conductivity.

As shown in Fig.2,
the effect of large leakage current, which is triggered by the largest voltage that
each sweeping would reaches at,
is more significant (compares to the polarization current) 
for the sweeping direction from large voltage toward zero.
Meanwhile the polarization-bound charge is hard to maintained due to the metallicity which
destroy the upward built-in electric field (by the mobile charge) in channel.
Such critical value for the dominating leakage current is weakly 
asymmetry with respect to the voltage, and results in different current response to the positive and negative voltage.
Furthermore,
the polarization in WTe$_{2}$ becomes more sensitive to the length of channel,
where the polarization field as well as the upward built-in electric field 
totally suppressed by the electric field along the channel,
in which case the delocalization of carrier plays the main role
and the polariation states can nearly no longer be found.

		\begin{figure}
				\centering
				\begin{subfigure}
					\centering
					\includegraphics[width=0.6\linewidth]{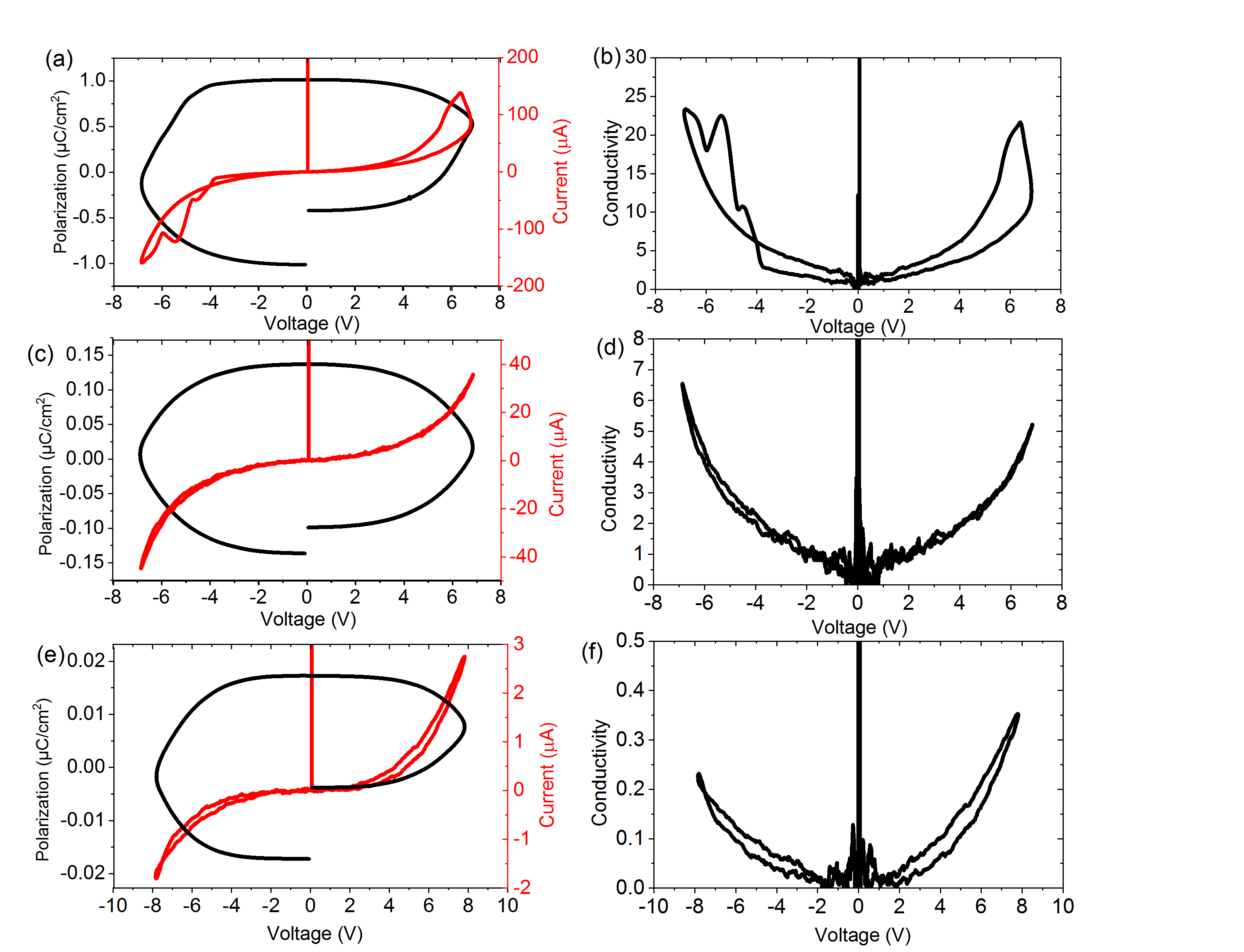}
				\end{subfigure}
		\captionsetup{font=small}
 	\caption{
P-V loop measurement of 10 nm WTe$_{2}$ thin film using DHM method.
				}
			\end{figure}

		\begin{figure}
				\centering
				\begin{subfigure}
					\centering
					\includegraphics[width=0.6\linewidth]{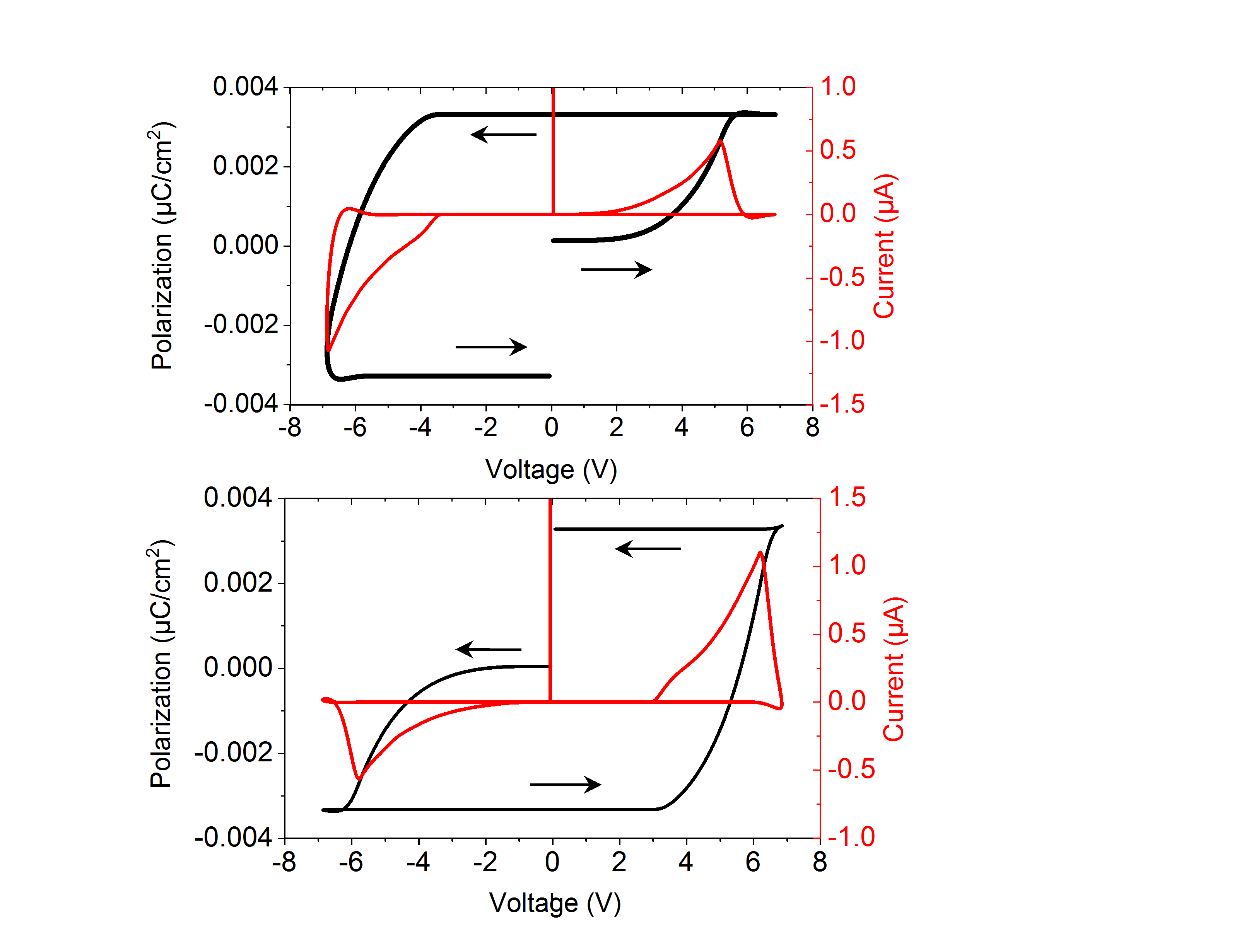}
				\end{subfigure}		\captionsetup{font=small}
				\caption{Ferroelectroc performance at low current realized by higher frequency.
With incresing voltage, there is a accumulation of voltage pulses that results in the strengthen
of ferroelectric polarization. But it is a short-term potentiation with volatile
polarization. The arrows indicate the sweeping direction.
				}
			\end{figure}

			{\color{red} $Conclusion$}
		There is significant role for the
		semimetal charge transport in
the related phenomenons, especially at nearly equal hole and electron densities.
		Except the experimental observation,
		we theortically consider the highly anisotropic electronic structure near the type-II Weyl cone of WTe$_{2}$ in terms of the low-energy two band model.
The calculation of dipole transition matrix element exhibit great different to other Dirac or Weyl (like the type-I) low-dimensional materials,
		which is due to its nearly compensated fermi surface and the anisotropic response to the magnetic field.
		Except the electrical anisotropicity as well as the outstanding out-of-plane ferroelectrcity,
		the calculation of dipole (modelled using a one-dimensional effective external fied) can also explain the special pattern of the resonance in semimetal charge transport which is different to that in other mterials with saturating magnetoresistivity\cite{Ali M N}.
	Considering the interband contributions,
			we calculation, in terms of dipole transition, the anisotropic electrical response to external field.			
			Hysteresis originates from ferroelectric polarization also observed through the hysteresis loop of the out-of-plane polarization.
			
			\clearpage
			
					{\color{red} {\bf Appendix}}\\
					\\
						{\color{red} $Analytical\ calculatons\ and numerical\ simulations$}

			In the presence of SOC,
			there are not Weyl nodes at $k_{z}\neq 0$\cite{na},
			and the tilting is along the $k_{y}$ direction ($b$-axis of WTe$_{2}$),
			which makes the chiral photocurrent response origining from different tilting of Weyl cones
			is impossible to exist for a normally incident light.
			Such a chiral photocurrent is possible only by a circularly polarized light which could induce spin-flipping
			and be irradiated along the tilting direction ($b$-axis),
			due to the intrinsic noncentrasymmetry properties
			where both the time-reversal symmetry and inversion symmetyr are broken.
Although the inversion symmetry is absence in Weyl semimetal WTe$_{2}$,
			the in-plane two-fold rotational symmetry is preserved\cite{Li P} as indicated by the mirror symmetry plane $\mathcal{M}_{yz}$
			and glide mirror symmetry plane $\mathcal{M}_{xz}$.
			This in-plane symmetry will leads to cancellation of any in-plane optical response induced by normally-incident circularly-polarized light.
			Thus the nonlinear AC photoconductivity tensors
			are vanishingly small other than the $\chi_{zxx}$ and $\chi_{zyy}$
			(which are directly related to the induced current in $z$-direction which does not affected by the twofold rotational symmetry),
			base on the experimentally obtained WTe$_{2}$ band structure near a Weyl cone.
			Such an in-plane responses cancellation will vanish for third-order Harmonic generation.
Similarly, the chiral anomaly-induced linear or nonlinear Hall current 
			is only possible to realized by a pair of non-orthogonal magnetic field and electric field\cite{Li P},
			applied in the direction of tilting.
			Thus a vertical magnetic field can still be applied to enhance the electron interactions. That is also why we consider only one Weyl node,
			as the internode scattering or the internode current (like chiral anomaly) are absent without the strong magnetic field or light field along $b$-aixs.
Another difference to the photocurrent process is that,
			for circularly polarized light-drived photocurrent,
			the excitation usually has a long relaxation time due to the large weight of optical absorption
			which may even increased for larger layer number and thus reduce the SHG intensity.
			The enhancement of circular photogalvanic effect,
			can effectively suppress the optical absorption by reducing the
			the relaxation time for a perturbed electron back to equilibrium.

			For materials lacking the inversion symmetry
			like MoS$_{2}$, hBN, and 1T$_{d}$ WTe$_{2}$,
			the nonlinear effects like the SHG or HHG are available without needing a material junction or bias voltage.
			As a low-symmetry two-dimensional semimetal,
			WTe$_{2}$ is also a promising candidate for the nonlinear effect,
			like the current-induced SHG through the applying a bias or gate voltage,
			and the circular photovoltage response.
			During these explorations,
			it is vital in keeping the equilibrium distribution of electrons.
			For example, by selecting a proper low-frequency of light (at near-infrared range and satisfying the optical resonance),
			we can greatly avoid the optical absorption as well as the off-resonance excitaions,
			which is a linear response
			and can be enhanced by increasing layer number of WTe$_{2}$
			and thus suppresses the SHG.
			Besides, the optical absorption is possibly leads to photodamages.
			For WTe$_{2}$ in$T_{d}$ phase which is noncentrosymmetric,
			its two-fold rotational symmetry with a mirror plane $\mathcal{M}_{yz}$ and a glide mirror plane $\mathcal{M}_{xz}$
			results in an in-plane inversion symmetry,
			and thus leads to the cancellation of of any in-plane components of photocurrent
			induced by a normally incident circularly polarized light.
			When considering the spin and Weyl cone tilting,
			there is actually a compound symmetry comprising both the time-reversal and two-fold rotation
			where the time-reversal process here has two effects: reverse the tilting direction and flip the spin.
			Experimental result\cite{Lim S} shows that for WTe$_{2}$ under obiquely incudent light, 
			the in-plane nonzero polarization-dependent current is only possible in the presence of a $z$ polarization,
			while the in-plane nonzero spin current is only possible 
			in the presence of nonzero electromagnetic potential components in three directions,
			i.e., the propagating direction of photon cannot be parallel to any crystal axises.
			Thus for a normally incident light, we can efficiently avoid the disturbations from in-plane photocurrents.
			As a response to the source-drain field variation, the ferroelectric polarization
			can be induced even by a dc electric current applied along the $a-b$ surface of WTe$_{2}$,
			and this can be understood as an excitation to a state that containing more out-of-plane (vertical) polarization,
			in addition to the spontaneous polarization due to the lack of inversion symmetry.
			Similar polarization effect can be seem from the magnetic-sweeping -direction dependence of the Hall resistivity at low-temperature limit.

				{\color{red} $Tight-binding\ model\ analysis$}
			
				For WTe$_{2}$,
		there are four pairs of Weyl points in the $k_{z}=0$ plane,
		and each containing two Weyl points with opposite chiralities\cite{na}.
		For most Dirac/Weyl systems, the energy
		dispersion is linear along the primary rotation axis, and be linear, higher in power
		for the in-plane directions.
		For Weyl system the power of in-plane dispersion corresponds to the account of degenerated conventional Weyl fermions with the same chirality.
		Specifically,
		to ensuring the Weyl point be of type-II,
		along the primary axis, 
		it must satisfies that the electron and hole pockets are touched and
		the kinetic energy is dominant compared the potential energy.
		Also, a vector-like potential behavior and the chiral zero mode
		(the so-called chiral anomaly) only appear when the
		direction of the in-plane magnetic field is within this range.
		For WTe$_{2}$, this direction is roughly along the direction of $b$-axis,
		where there is a Weyl nodal line connecting two Weyl points with opposite chiralities.
		We use the location of one of the in-plane Weyl points reported in Ref.\cite{na},
		which is $(k_{x},k_{y})=(0.1241, 0.0454)$ eV$\r{A}$.
		While another Weyl point is at $(k_{x},k_{y})=(0.12184, 0.03825)$,
		then the direction of magnetic field validating the chiral anomaly
		$\tan\theta=\frac{k_{y}}{k_{x}}\le -3.8$ or $\tan\theta=\frac{k_{y}}{k_{x}}\ge 0.55$.
		According to the experiments conducted in this work,
		we further explore the different effects brought by the magnetic field or electric current
		along different in-plane directions of the WTe$_{2}$ bulk crystal sample in 
		$T_{d}$ phase.
		This also reveals the anisotropic structure of the WTe$_{2}$.
		Due to the anisotropic fermi surface, which results in the magnetic field selection for the chiral anomaly,
		WTe$_{2}$ displays a different degree of non-saturating
		magnetoresistivity and electron-hole compensation.
		Along $b$-axis which is the domained by the kinetic term and just within the Weyl cone,
		the effect of compensated fermi surfaces as well as the Weyl electronic structure
		is much lighter than that in the other directions.
		While the electron-hole compensation is essential in non-saturating magnetoresistivity,
		the direction where chiral anomaly could be found may exhibit stronger dependence on the multiband effect (larger overlap between the valence band and conduction band).

			For general Weyl semimetals,
			we consider the following Hamiltonian
			\begin{equation} 
				\begin{aligned}
					H=\chi \varepsilon_{0}(k_{x}\sigma_{x}+k_{y}\sigma_{y}+k_{z}\sigma_{z})-\mu\sigma_{0},
				\end{aligned}
			\end{equation}
			where in a lattice model $\varepsilon_{0}=\hbar v=ta$ where $t$ is the hopping integral and $a$ is the lattice constant.
			For such linearized Hamiltonian,
			the generalized derivatives $v_{mn}^{ab}$ can be ignored.
			For convenience in calculation,
			we use the following coordinate transformation
			\begin{equation} 
				\begin{aligned}
					&k_{x}=\rho {\rm sin}\theta {\rm cos}\phi,\\
					&k_{y}=\rho {\rm sin}\theta {\rm sin}\phi,\\
					&k_{z}=\rho {\rm cos}\theta.
				\end{aligned}
			\end{equation}
			Thus we have
			\begin{equation} 
				\begin{aligned}
					dk_{x}dk_{y}dk_{z}=\rho^{2}{\rm sin}\theta d\rho d\theta d\phi,
				\end{aligned}
			\end{equation}
			and the eigenvalues of the above Hamiltonian reads $\varepsilon_{m=0,1}=-\mu\pm \chi\rho$
			for a two-band model.
			Then the velocity operators $v^{a}$ can be obtained as
			\begin{equation} 
				\begin{aligned}
					&v^{x}=\chi
					\begin{pmatrix}
						{\rm cos}2\theta {\rm csc}\theta {\rm sec}\theta {\rm sec}\phi & (1+i{\rm cot}\phi)(-2i+{\rm cot}\phi){\rm tan}\phi\\
						(1-i{\rm cot}\phi)(2i+{\rm cot}\phi){\rm tan}\phi & -{\rm cos}2\theta {\rm csc}\theta {\rm sec}\theta {\rm sec}\phi
					\end{pmatrix},\\
					&v^{y}=\chi
					\begin{pmatrix}
						{\rm cos}2\theta {\rm csc}\theta {\rm csc}\phi {\rm sec}\theta & -3i+2{\rm cot}\phi-{\rm tan}\phi\\
						3i+2{\rm cot}\phi-{\rm tan}\phi & -{\rm cos}2\theta {\rm csc}\theta {\rm csc}\phi {\rm sec}\theta
					\end{pmatrix},\\
					&v^{z}=\chi
					\begin{pmatrix}
						2 & {\rm cos}2\theta {\rm csc}\theta {\rm sec}\theta (-{\rm cos}\phi+i{\rm sin}\phi)\\
						{\rm cos}2\theta {\rm csc}\theta {\rm sec}\theta (-{\rm cos}\phi-i{\rm sin}\phi) & -2
					\end{pmatrix},
				\end{aligned}
			\end{equation}
			and the matrix elements can be obtained as
			\begin{equation} 
				\begin{aligned}
					v_{mn}^{x}
					&=\langle m|v^{x}|n\rangle\\
					&=\delta_{mn}\langle v^{x}\rangle (\frac{k_{x}}{\rho})(-\delta_{m,0}+\delta_{m,1})
					+(1-\delta_{mn})\langle v^{x}\rangle (\frac{k_{z}}{\rho}\frac{k_{x}}{\sqrt{k_{x}^{2}+k_{y}^{2}}}-i\frac{k_{y}}{\sqrt{k_{x}^{2}+k_{y}^{2}}})\\
					&=\delta_{mn}\langle v^{x}\rangle ({\rm sin}\theta{\rm cos}\phi)(-\delta_{m,0}+\delta_{m,1})
					+(1-\delta_{mn})\langle v^{x}\rangle ({\rm cos}\theta{\rm cos}\phi-i{\rm sin}\phi),\\
					v_{mn}^{y}
					&=\delta_{mn}\langle v^{y}\rangle ({\rm sin}\theta{\rm sin}\phi)(-\delta_{m,0}+\delta_{m,1})
					+(1-\delta_{mn})\langle v^{y}\rangle ({\rm cos}\theta{\rm sin}\phi-i{\rm cos}\phi),\\
					v_{mn}^{z}
					&=\delta_{mn}\langle v^{z}\rangle {\rm cos}\theta(-\delta_{m,0}+\delta_{m,1})
					+(1-\delta_{mn})\langle v^{z}\rangle \frac{-{\rm sin}\theta}{\rho},\\
				\end{aligned}
			\end{equation}
			where the additional term $(-\delta_{m,0}+\delta_{m,1})$ consider the two-band model
			in which case the intraband velocity depends on the band index $m,n=0,1$.
			$\langle v^{a}\rangle$ denotes the eigenvalues of matrix $v^{a}$.

			When the effect of spin-orbit coupling
			is considered,
			there are gapless Weyl nodes in type-II Weyl semimetal WTe$_{2}$.
			Similar to graphene,
			when the electron move from
			the linear valence band to the linear conduction band near a Weyl node,
			or vice versa,
			there will be a strong instantaneous accelaration when the 
			electron close the node,
			and leads to strong emission of detectable radiation in the mean time.
			That means the electrical dipole effect considered during this process 
			mostly comes from the interband transition,
			the electron will turning back to the equilibrium position after the perturbation.
			But the broden peak at finite temperature implies the existence of relaxations,
			due to, e.g., the impurity or phonon scatterings.

			For WTe$_{2}$ in $Pmn2_{1}$ space group,
			both the time-reversla symmetry and the inversion symmetry are broken,
			which give rise to nonzero Berry curvature.
			When the spin-orbit coupling (SOC) is being considered
			to make the interband transition is possible\cite{na},
			the layer number indeed affects littlely on the band structure as well as the valence band SOC splitting\cite{Liu G B}
			of WTe$_{2}$.
			Next we using the experimentally obtained parameters of Ref.\cite{na},
			where the Hamiltonian of type-II Weyl semimetal WTe$_{2}$ reads
			\begin{equation} 
				\begin{aligned}
					H=Ak_{x}\sigma_{0}+Bk_{y}\sigma_{0}+ak_{x}\sigma_{y}+bk_{x}\sigma_{z}+dk_{y}\sigma_{z}+ek_{z}\sigma_{x},
				\end{aligned}
			\end{equation}
			with $A=-2.7,B=0.6,a=1,b=1.1,d=0.27,e=0.184$.
			Following the above coordinate transformation,
			the matrix form of the Hamiltonian is
			\begin{equation} 
				\begin{aligned}
					H=
					\begin{pmatrix}
						(A\rho+b\rho){\rm cos}\phi{\rm sin}\theta+(B\rho+d\rho){\rm sin}\phi {\rm sin}\theta & e\rho{\rm cos}\theta-ia\rho{\rm cos}\phi{\rm sin}\theta\\
						e\rho{\rm cos}\theta+ia\rho{\rm cos}\phi{\rm sin}\theta & (A\rho-b\rho){\rm cos}\phi{\rm sin}\theta+(B\rho-d\rho){\rm sin}\phi {\rm sin}\theta 
					\end{pmatrix}.
				\end{aligned}
			\end{equation}
			After substituding the above parameters,
			we obtain the eigenvalue of the Hamiltonian as $\varepsilon_{+}=1.06\rho\hbar v_{0}$ for the conduction band,
			and $\varepsilon_{-}=0.7\rho\hbar v_{0}$ for valence band.
			Here we select the points at $|k|=ta$ away from the Weyl node,
			and setting the chemical potential as $\mu\approx (\varepsilon_{+}+\varepsilon_{-})/2$
			to prevent the optical absorption when the incoming photon energies is higher than the band gap.
			Thus for WTe$_{2}$, by turning the chemical potential to $\mu=\rho\hbar v_{0}$,
			we redefinen the energies of the three levels as $\varepsilon_{m}=-0.3\rho\hbar v_{0},
			\varepsilon_{n}=-0.15\rho\hbar v_{0},\ \varepsilon_{l}=1.06\rho\hbar v_{0}$.
			Both of these two energies are positive,
			indicating that one of them is electron band while the other is hole band,
			with the same topologically protected chirality.
			
			Since the type-II Weyl point requires kinetic energy dominates\cite{na},
			and thus $\phi$ should be in the range $0.58\pi>\phi>\pi/2$ or $1.58\pi>\phi>3\pi/2$.
			In this case, through the same procedure shown in above and by taking $\phi=0.55\pi$,
			we obtain the eigenvalues of velocity operators as
			$\langle v_{x}\rangle\approx -12.3\varepsilon_{0}/\hbar$, $\langle v_{y}\rangle\approx -4.9\varepsilon_{0}/\hbar$
			and $\langle v_{z}\rangle\approx 0.75\varepsilon_{0}/\hbar$.

			{\color{red} $Electrical\ dipole\ transition\ moment\ in\ two \ level\ approximation$}

			The ferroelectric switching as well as the polarization for
			low-symmetry ferroelectric material is closely related to the 
			electrically tunable spontaneous dipole,
			like the long range electrostatic interaction
			and the surface dipoles which can be modulated 
			by the interlayer shear displacement\cite{Ni Z}.
			In this section,
			we using the two level approximation in the semiclassical framework 
			to discuss the electrical dipole in bulk WTe$_{2}$ as well as its effect on the ferroelectric polarization.
			The dipole transition provides a way to study the low-energy
			excitations around the type-II Weyl point,
			in terms of a two-level approximation.
			We mainly consider the dipole along the $k_{y}$-direction
			in the $k_{z}=0$ plane,
			where there is a part of Chern number with opposite sign.
			In term our theortical model,
			the possibility to modify the electron-hole correlation near the Fermi surface (contains the electron and hole pocket) by
			applying an in-plane electric field is further confirmed.
			
			The perfect charge compensation (comparable electron and hole concentrations) in WTe$_{2}$ allows the usage of two-level approximation here.
			According to the parameter in Ref.\cite{na},
			we consider one of the Weyl cone in $x-y$ plane within the first Brillouin zone,
			where the relativistic Weyl fermions are described by the dispersion
			\begin{equation} 
				\begin{aligned}
					\varepsilon_{\pm}=1.204 k_{x}+0.686 k_{y}\pm
					\sqrt{(-1.159 k_{x})^2+(1.046 k_{x}+0.055 k_{y})^2},
				\end{aligned}
			\end{equation}
			where the Weyl point is locates in $(k_{x},k_{y})=
			(0.1241, 0.0454)$,
			and here a 
			local stable region near the Weyl node with a point-like degeneracy is resctricted 
			by the range $k_{x}^2+k_{y}^2=10^{-6}$\cite{na}.
			As a superposition of the superposition of eigenstates,
			the wavefunction reads
			\begin{equation} 
				\begin{aligned}
					\Psi(t,r)=\frac{1}{N}\sum_{k\in BZ}(\sum_{n}u_{n}(k ,{\bf r})\phi_{n}(k,t)
					+\sum_{m}u_{m}(k ,{\bf r})\phi_{m}(k,t)),
				\end{aligned}
			\end{equation}
			where $u_{n}(k ,{\bf r})$ and $u_{m}(k ,{\bf r})$ are the periodic part of the Bloch wave functions.
			
			\begin{figure}
				\centering
				\begin{subfigure}
					\centering
					\includegraphics[width=0.4\linewidth]{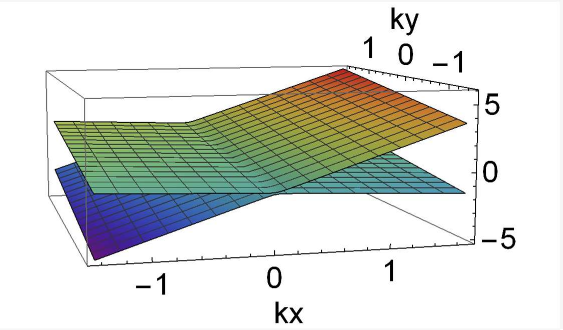}
				\end{subfigure}
				\caption{Band structure of WTe$_{2}$ near the Weyl node located at $(k_{x},k_{y})=$(0.1241, 0.0454).
					We show the region of $(0.1241\pm\pi/2 , 0.0454\pm\pi/2)$ in momentum space.
				}
			\end{figure}

			\begin{figure}
				\centering
				\begin{subfigure}
					\centering
					\includegraphics[width=0.4\linewidth]{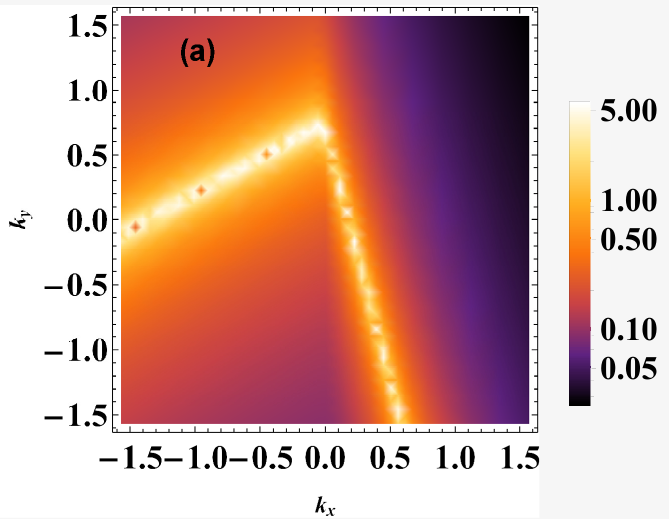}
				\end{subfigure}
				\begin{subfigure}
					\centering
					\includegraphics[width=0.4\linewidth]{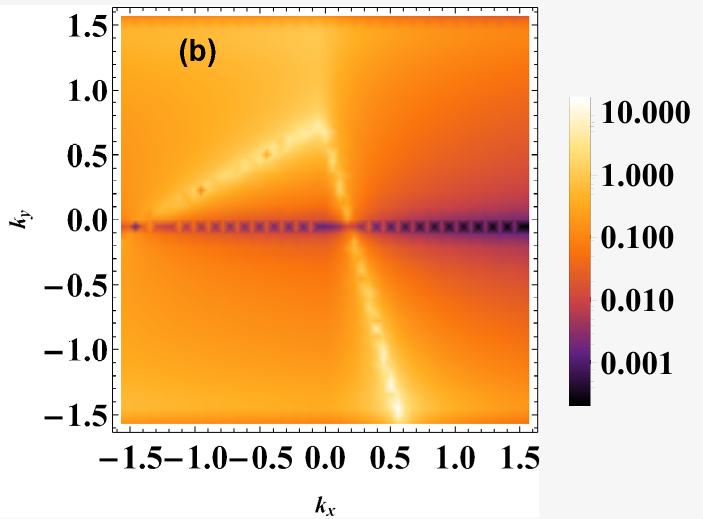}
				\end{subfigure}
				\caption{Real part (a) and imaginary part (b) of the interband dipole with electric field at frequency $\omega=0$.
				}\label{12}
			\end{figure}
			
			\begin{figure}
				\centering
				\begin{subfigure}
					\centering
					\includegraphics[width=0.4\linewidth]{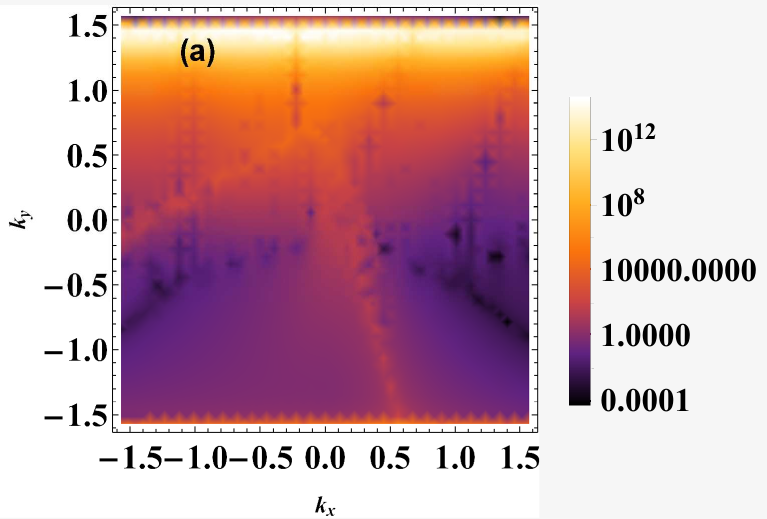}
				\end{subfigure}
				\begin{subfigure}
					\centering
					\includegraphics[width=0.4\linewidth]{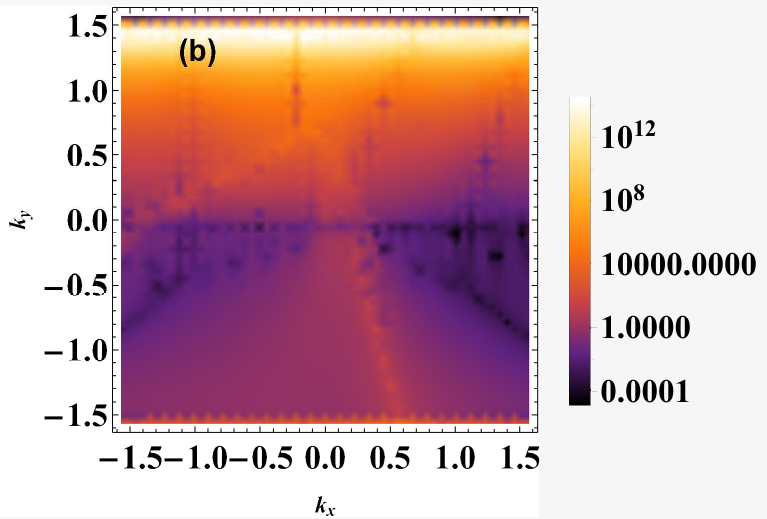}
				\end{subfigure}
				\caption{The same with Fig.\ref{12} but with field frequency $\omega=\pi$.
				}
			\end{figure}

			For arbitary $\alpha$, we have
			${\rm cos}\theta_{k}=e^{2 i {\bf kr}}-1$ which directly determines the energy dispersion,
			thus the in the dipole transition process, the transition frequency reads
			$|\varepsilon_{n}-\varepsilon_{m}|=\Delta_{0} {\rm cos}^{-1}\theta_{k}$
			where $\Delta_{0}$ is the momentum-independent free electron energy
			(the free particle kinetic energy which can be viewed as a constant here),
			and considering the classical saddle point analysis,
			the effective action can be obtained from the transition frequency in time domain as
			$S(k ,t,t')=\int^{t}_{t'}d\tau|\varepsilon_{n}-\varepsilon_{m}|(\tau)$.
			In terms of the (interband) transition dipole,
			we consider the hopping integral as $
					f(k )=
					-2\cos k_{x}-2\cos k_{y}+2i\sin k_{x}+2i\sin k_{y},
$,
			and the dipole matrix element reads $
					D(k )=\langle u_{n}(k ,{\bf r})|{\bf r}|u_{m}(k ,{\bf r})\rangle=\frac{1}{2}
					\partial_{k }\psi(k )$.
			The non-adiabatic type coupling can be regarded as originates from the dynamics of the mixing angle,
			i.e., $\dot{\theta}_{k}$.
			Before substitution,
			the set of TDSE reads
			\begin{equation} 
				\begin{aligned}
					E\phi_{n}(k )
					=&(E_{n}(k )-E(t)D_{nm}-iE(t)\partial_{k})\phi_{n}(k )
					-\frac{E(t)}{2}\partial_{k}\psi_{k}\phi_{m}(k ),\\
					E\phi_{m}(k )
					=&(E_{m}(k )-E(t)D_{nm}-iE(t) \partial_{k})\phi_{m}(k )
					-\frac{E(t)}{2}\partial_{k}\psi_{k}\phi_{n}(k ),
				\end{aligned}
			\end{equation} 
			Considering the linearly polarized electric field, we have
			\begin{equation} 
				\begin{aligned}
					E\phi_{n}(k )
					=&(E_{n}(k )- E (t)D_{x}(k_{y})-i E (t)\frac{\partial}{\partial k_{x}})\phi_{n}(k )
					-\frac{ E (t)}{2}\partial_{k_{x}}\psi_{k }\phi_{m}(k ),\\
					E\phi_{m}(k )
					=&(E_{m}(k )- E (t)D_{x}(k_{y})-i E (t)\frac{\partial}{\partial k_{x}})\phi_{m}(k )
					-\frac{ E (t)}{2}\partial_{k_{x}}\psi_{k }\phi_{n}(k ),\\
					E\phi_{n}(k )
					=&(E_{n}(k )- E (t)D_{y}(k_{x})-i E (t)\frac{\partial}{\partial k_{y}})\phi_{n}(k )
					-\frac{ E (t)}{2}\partial_{k_{y}}\psi_{k }\phi_{m}(k ),\\
					E\phi_{m}(k )
					=&(E_{m}(k )- E (t)D_{y}(k_{x})-i E (t)\frac{\partial}{\partial k_{y}})\phi_{m}(k )
					-\frac{ E (t)}{2}\partial_{k_{y}}\psi_{k }\phi_{n}(k ),
				\end{aligned}
			\end{equation} 
with the solved coefficients
			\begin{equation} 
				\begin{aligned}
					\label{coe1}
					\phi_{n}(k_{x})
					=&A_{1}{\rm Exp}\left[-\frac{i}{ E (t)}[Ek_{x}-\int^{k_{x}}_{0}[E_{n}- E (t)D_{x}(k_{y})]dk_{y}]\right],\\
					\phi_{m}(k_{x})
					=&A_{2}{\rm Exp}\left[-\frac{i}{ E (t)}[Ek_{x}-\int^{k_{x}}_{0}[E_{m}- E (t)D_{x}(k_{y})]dk_{y}]\right],\\
					\phi_{n}(k_{y})
					=&A_{3}{\rm Exp}\left[-\frac{i}{ E (t)}[Ek_{y}-\int^{k_{y}}_{0}[E_{n}- E (t)D_{y}(k_{x})]dk_{x}]\right],\\
					\phi_{m}(k_{y})
					=&A_{4}{\rm Exp}\left[-\frac{i}{ E (t)}[Ek_{y}-\int^{k_{y}}_{0}[E_{m}- E (t)D_{y}(k_{x})]dk_{x}]\right].
				\end{aligned}
			\end{equation} 
			It shows that only the $\phi_{n}(k_{x})$ and $\phi_{m}(k_{y})$ be the reasonal results due to the special form of 1D dipoles $D_{x}(k_{y})$ and $D_{y}(k_{x})$.
			Using the dipole transition matrix elements
			and according to Larmor semiclassical theory,the low-energy
			excitation around type-II Weyl point due to the dipole acceleration
			can be obtained by considering the following interband contribution $
					{\bf d}(t)=\int dk D(k )(\phi_{n}(k,t)\phi_{m}^{*}(k,t)+h.c.)$.

			\renewcommand\refname{References}
			
			\clearpage

		\end{small}
	\end{document}